\documentclass[superscriptaddress,twocolumn,showpacs,preprintnumbers,prb]{revtex4}
\usepackage{graphicx}
\usepackage{times}
\usepackage{epsfig}
\begin{document}

\def\salto{\vskip 1cm}
\def\lag{\langle}
\def\rag{\rangle}

\newcommand{\UT} {Condensed Matter Sciences Division, Oak Ridge National Laboratory, Oak Ridge, 
TN 37831 and Department of Physics, The University of Tennessee, Knoxville, TN 37996}
\newcommand{\FSU}{National High Magnetic Field Laboratory and Department of Physics, Florida 
State University, Tallahassee, FL 32306}
\newcommand{\UCSB} {Microsoft Project Q, The University of California at Santa Barbara, 
Santa Barbara, CA 93106}
\newcommand{\Rosario} {Universidad Nacional de Rosario, Avenida Pellegrini 250, 2000-Rosario, Argentina}

\title{Applying Adaptive Time-Dependent DMRG to Calculate \\
the Conductance of Strongly Correlated Nanostructures}

\author{K.A. Al-Hassanieh }\affiliation {\UT}\affiliation {\FSU} 
\author{A.E. Feiguin }     \affiliation {\UCSB}
\author{J.A. Riera}        \affiliation {\Rosario}
\author{C.A. B\"usser}     \affiliation {\UT}
\author{E.   Dagotto }     \affiliation {\UT}

\begin{abstract}
A procedure based on the recently developed ``adaptive'' time-dependent density-matrix-renormalization-group
(DMRG) technique is presented to calculate the zero temperature conductance of nanostructures, 
such as a quantum dots (QD's) or molecular conductors,
when represented by a small number of active levels.  The leads are modeled using 
non-interacting tight-binding Hamiltonians.  The ground state at time zero is calculated at zero bias.  Then, a small bias 
is applied between the two leads, the wave-function is DMRG evolved in time, and currents are measured as 
a function of time.  Typically, the current is expected to present
 periodicities over long times, involving intermediate 
well-defined plateaus that resemble steady states. The conductance can be obtained from those steady-state-like 
currents.  To test this approach, several cases of interacting and non-interacting systems have been studied.  Our 
results show excellent agreement with exact results in the non-interacting case.  More importantly, the technique
also reproduces quantitatively well-established results for the conductance and local density-of-states in both the 
cases of one and two coupled interacting QD's. The technique also works at finite
bias voltages, and it can be extended to include interactions in the leads.
\end{abstract}

\pacs{73.63.-b,71.27.+a,72.10-d,85.65.+h}
\maketitle

\section{INTRODUCTION}

The rapidly developing investigations in the area of nanometer-scale systems
and its concomitant potential technological applications in real devices
have induced considerable interest in the study of electrical transport
through small molecules and quantum dots. In fact, the construction of molecular
electronic devices \cite{ratner,molecular} is among the most exciting
areas of investigations in physics, and theoretical guidance is needed for the
success of this vast effort. Molecules can change their shape and position
relative to the leads as electrons enter 
or leave the molecule, making
the study of these systems very challenging. Moreover, 
Coulomb correlations cannot be
neglected in small devices. For a conceptual understanding
of these complex systems, it is imperative to develop models and
unbiased many-body methods that rely on a minimal number of assumptions, in order
to accurately handle both strong Coulombic and electron-phonon couplings. 
Quantum dots constructed using conventional semiconductor
technology also provide an important playground for the analysis of
transport properties in nanoscopic systems, and the theoretical challenges
in this context are equally important \cite{quantum-dots}.

The conductance of small nanoscopic systems can be theoretically
estimated using a variety of techniques. Among the most popular approaches
are the ab-initio calculations using density functional theory (DFT).
These one-electron self-consistent methods have been successful in
describing various I-V characteristics \cite{DFT1,DFT2,DFT3}. 
However, the applicability of
these ideas must be carefully scrutinized, as recently remarked
by Muralidharan {\it et al.} \cite{datta}. For example, 
it is clear that in small molecules
charging effects are important, and they effectively act as quantum dots
in the Coulomb Blockade regime. Moreover, techniques that do not take
into account the strong correlation between electrons cannot capture
important effects such as the Kondo resonance (arising from the
coupling between localized spins and conduction electrons) which induces a new
channel for transport in a variety of small systems \cite{Kondo-experiment,Kondo-theory}. 
In addition,
it is well known that several bulk materials, such as transition metal oxides,
cannot be described with ab-initio methods that neglect correlations.
The complexity of their behavior, including potentially useful
effects such as large magnetoresistances in Mn oxides \cite{Dagotto-review}, may manifest
in nanoscopic systems as well, and the use of strongly correlated
materials in nanodevices may lead to interesting applications. To
study all these systems (small molecules, quantum dots, and in general
nanodevices that include strongly correlated materials), techniques
beyond DFT must be developed. 

A similar challenge occurred before in the study of bulk materials and several
years of research have shown that the following 
two-steps process leads to profound
insights. The first step consists of a simple modeling of the material,
typically either deducing the relevant degrees of freedom from atomistic
considerations when the states are very localized, or borrowing from
band structure calculations to isolate the minimal ingredients needed
to capture the essence of the problem.
The second step, the hardest, is solving the resulting model, which is typically
of a tight-binding nature with the addition of Coulombic and phononic couplings. In the
strong coupling regime, the use of numerical techniques provides the
most reliable unbiased approach for the approximate investigation 
of tight-binding-like models with Coulombic interactions \cite{RMP}. As a consequence,
a natural path toward the study of transport in strongly correlated
nanoscopic systems can also start with simple models and use computational
techniques for their analysis. It is the main purpose of this paper
to propose a technique that can be used to study transport in
systems described by strongly correlated electronic models.

The method to calculate conductances 
proposed here relies on the successful Density Matrix
Renormalization Group (DMRG) technique \cite{White-1992,RMPSch}. While in principle
this  method is applicable only
to quasi-one-dimensional models, such a geometry is quite acceptable
in several nanoscopic important problems including transport in arrays of quantum dots
or using small molecules as bridges between leads. However, a straightforward
application of the original DMRG methodology is not immediately useful
to study transport, particularly when arbitrarily large 
external electric and magnetic fields, with potentially complicated time
dependences, are switched on and off at particular times. Nevertheless, progress toward a
computational tool for these type of problems has been steady in recent years.
For instance,
important numerical methods for dynamical DMRG studies were presented
to handle frequency dependent spectral functions \cite{karen}. More directly
focusing on real-time investigations, interesting techniques were 
proposed \cite{marston}. While useful for many qualitative applications, 
these methods are in general not
as accurate and stable as needed for the detailed study of finite bias 
transport in complicated nanosystems. The reason is that the method of Ref.\onlinecite{marston}
is ``static'' in the sense that the truncated Hilbert space found to be optimal
at time $t$=0, namely before switching on the external fields, is kept at all
times. This approach breaks down after relatively short times, since extra states
are needed for a proper description of transport at finite times. A further approximation
to improve on this first proposal
is to enlarge substantially the initial Hilbert space so that it remains suitable for properties
calculated at finite times \cite{Luo}. 
This technique has the problem that the number of
states grows rapidly with the simulation time and eventually it becomes impractical.
Nevertheless, the method has been successfully used to study the propagation of
a density excitation in an interacting clean system  \cite{Schmitteckert1}.

Recently, important developments have led to the ``adaptive'' time-dependent version
of the DMRG method, that is efficient over long times and, thus, it is
suitable to handle the problems 
we are focusing on (for a detailed review see Ref.\onlinecite{Sch}). The method
to be used here was developed independently by White and Feiguin \cite{White} 
and Daley {\it et al.} \cite{Daley}, after the idea of how to do time-evolution
to a matrix product was introduced by Vidal \cite{Vidal}, 
and relies on an adaptive optimal Hilbert space that follows the
state as time grows. The method is based on a 
Suzuki-Trotter break-up of the evolution operator, and as a consequence 
a Trotter truncation error is introduced. Fortunately, this systematic error
can be easily estimated and controlled. The adaptive DMRG numerical method will only
be briefly reviewed below since our proposal uses the technique to calculate
conductances, but does not modify the method itself.
The reader should consult the original literature \cite{White,Daley}
for more details. It is important to remark that the technique
is easy to implement once a ground-state DMRG code is prepared and, moreover,
the time-evolution is stable, as shown explicitly in our results and in some previous
investigations (further
improvements can be added with the time-step targeting method recently proposed
by Feiguin and White \cite{Feiguin}).
The conclusion of our effort documented below 
is that the adaptive method provides accurate results for
the calculation of conductances. The technique has passed the test of
noninteracting electrons, as well as the cases of one and two interacting quantum
dots, where a subtle Kondo effect occurs. Moreover, the method is not restricted to 
small biases but it produces reasonable answers at finite bias as well. As
a consequence, it has the potential of being the method of choice to study
transport under both weak and strong external fields, 
in small nanostructures of substantial complexity. 
Multilevel model Hamiltonians, possibly inspired by ab-initio calculations,
can be used to describe the ``bridge'' between leads. In addition, the method is
particularly transparent since it relies on the straightforward 
calculation of a current in the
presence of a voltage, rather than relying on other indirect linear-response formulas.

Of course, the reader must be aware that the
method is not of unlimited applicability. If the
molecules or Kondo clouds are too long in size, eventually not even DMRG can handle the
very long chains needed for a proper description. For completeness, and to assure a balanced
description of the technique, one of these difficult cases is also presented in our
manuscript. But often the qualitative physics
can be understood by relaxing parameters, thus we expect that even in very complicated 
cases the propose technique will be helpful, at least at the conceptual level. 
Other limitations of the present technique is that energy
dissipation is not incorporated, and the temperature is restricted to be zero. 
Improving on these issues is a task left for the near future.

It is important to remark that there are other numerical techniques that
can also be used to study transport in strongly correlated nano-systems. One of
them is the Numerical Renormalization Group, that evolved from Wilson's original
RG ideas. This technique is quite accurate, as exemplified by some recent
calculations \cite{NRG}, but it cannot be used for arbitrary
problems. Since our goal is to try to develop a method that can handle the fairly
complex models that will be used in the near future to represent, e.g., small
molecular conductors, this method does not have sufficient flexibility for our
purposes. In cases where NRG works, it should be the method of choice, but this
occurs in a small subset of problems in the area of transport across correlated systems.
A second approach relies on the static DMRG method, 
using a ring geometry and with a current induced by a flux threading the ring \cite{molina}.
A recently proposed third method combines linear response Kubo theory with static DMRG
and the conductance is calculated based on correlation functions
in the ground state \cite{Schmitteckert2}.
It would be interesting to find out if the methods of Refs.\onlinecite{molina,Schmitteckert2} 
can reach a similar accuracy as ours for the case of the one and two quantum dots.
A fourth method is the Exact Diagonalization technique followed by a Dyson equation
embedding procedure (ED+DE) \cite{ED+DE1,Torio} 
where the interacting region is solved exactly (including
some sites of the leads), and then the rest of the leads are taken into account via a Dyson
self-consistent approach. The method directly treats bulk systems, contrary to the DMRG technique that is
necessarily limited to a large but finite chain, 
it is flexible and has led to interesting results for
difficult cases, such as center-of-mass phonons in molecular conductors,
and multilevel systems \cite{Busser1,Busser2,Martins1,Martins2,Khaled}.
However, the Dyson embedding is somewhat arbitrary and it is 
difficult to control its accuracy. It is our intention in the near future to
combine the ED+DE method with the DMRG approach discussed in this paper, and for the
physical problems where these independent techniques give similar results, then the case can
be made that a reliable conclusion has been reached. Thus, ED+DE and the present method
are complementary.

The organization of the manuscript is as follows. After the present introduction,
in Section II, the models are defined and the technique is very briefly described.
Section III contains the important test of noninteracting electrons (note that although the
Coulombic coupling is zero, there are different hopping amplitudes at different links).
Here, the systematic behavior of the method is discussed in detail. Section IV deals
with the case of a quantum dot, with a nonzero Hubbard coupling. The value of $U$ is 
comparable to the hoppings, to prevent the Kondo cloud from reaching huge sizes that
would render the DMRG method useless (nevertheless, one ``large'' $U$ case is studied
for completeness, to illustrate the limitations of the method). Section V contains
the case of two dots, which themselves can be coupled in a ``singlet'' preventing conduction
or loosely coupled having individually a Kondo effect. The conclusions in Section VI
briefly summarize our findings.

\section{MODEL AND CONDUCTANCE CALCULATION}

\begin{figure}
\vspace{0.5cm}
\epsfxsize=8cm \centerline{\epsfbox{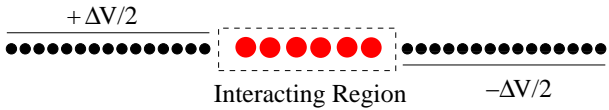}}
\caption{Schematic representation of the geometry used in our study.  The leads are modeled by tight-binding Hamiltonians.  
The ground state at time zero is calculated at zero bias. Then, a finite 
bias $\Delta V$ is applied between the two leads (without ramping time for simplicity, but this
can be changed in future studies)
and the resulting current is measured.}
\label{schematic}
\vspace{0.25cm}
\end{figure}

In general, the systems studied here consist of a relatively small region where Coulomb
interactions are present, weakly coupled to two non-interacting leads (see Fig.\ref{schematic}).  
The interacting region can represent one or several quantum dots (QD's), 
a single-molecule conductor, or other nanoscopic regions.  In fact, the generality of the method
presented in this paper, allows for a wide variety of interacting systems.

The leads are modeled as ideal tight-binding chains.  
As examples, the focus will be on one QD and two QD's connected 
in series.  The total Hamiltonian of theses systems can be written in general as   
\begin{equation}
\hat H = \hat H_{\rm leads} + \hat H_{\rm cluster} + \hat H_{\rm cluster-leads},
\label{Htotal} 
\end{equation}
where $\hat H_{\rm leads}$ is the Hamiltonian of the leads, which is
\begin{equation}
\hat H_{\rm leads} = -t_{\rm leads}\sum_{i\sigma} [c_{li\sigma}^{\dagger}c_{li+1\sigma} + c_{ri\sigma}^{\dagger}c_{ri+1\sigma} + h.c.].
\label{Hleads} 
\end{equation}
$t_{\rm leads}$ is the hopping amplitude in the leads, which in 
the following is taken as the 
energy scale (i.e. $t_{\rm leads}$ = 1).  The operator 
$c_{li\sigma}^\dagger$ ($c_{ri\sigma}^\dagger$) creates an electron 
with spin $\sigma$ at site $i$ in the left (right) lead.  $\hat H_{\rm cluster}$ is the Hamiltonian of the 
cluster where the interactions are present. 
Finally,  $\hat H_{\rm cluster-leads}$ is the Hamiltonian that connects the interacting region 
to the leads, typically a hopping term.

\subsection{One Quantum Dot}
For the case of one QD, represented simply by one active level, $\hat H_{\rm cluster}$ can be written as
\begin{equation}
\hat H_{\rm cluster} = V_{g} n_{d} + Un_{d\uparrow}n_{d\downarrow},
\label{Hdot} 
\end{equation}
where the first term represents the location of the 
energy level of the QD controlled by the gate voltage $V_{\rm g}$.  The second 
term represents the Hubbard repulsion between electrons of opposite spins occupying the QD.  
$n_d =n_{d\uparrow} + n_{d\downarrow}$ is the number of electrons at the dot.
$\hat H_{\rm cluster-leads}$ can be written as
\begin{equation}
\hat H_{\rm c-leads} = -t'\sum_{\sigma} [c_{l\sigma}^{\dagger}c_{d\sigma} + c_{r\sigma}^{\dagger}c_{d\sigma} + h.c.], 
\label{Hdotleads} 
\end{equation}
where $t'$ is the amplitude for the electronic hopping between the QD and the leads.  
$c_{d\sigma}^\dagger$ creates and electron with spin 
$\sigma$ at the dot, while  $c_{l\sigma}^\dagger$ creates an electron at the last site of the left lead and 
$c_{r\sigma}^\dagger$ creates an electron at the first site of the right lead, if sites are numbered from
left to right.

\subsection{Two Coupled Quantum Dots in Series}
In the case of two QD's, $\hat H_{\rm cluster}$ can be written as
\begin{equation}
\hat H_{\rm cluster} = \sum_{\alpha=1,2} [ V_{\rm g} n_{d\alpha} + Un_{d\alpha\uparrow}n_{d\alpha\downarrow}] - t'' \sum_{\sigma}[c_{d1\sigma}^\dagger c_{d2\sigma} + h.c.],
\label{Hdots} 
\end{equation}
where $n_{d\alpha} = n_{d\alpha\uparrow} + n_{d\alpha\downarrow}$ is the number of electrons at the quantum 
dot $\alpha$, 
and $t''$ is the hopping between the two dots.  $\hat H_{\rm cluster-leads}$ is written as
\begin{equation}
\hat H_{\rm cluster-leads} = -t'\sum_{\sigma} 
[c_{l\sigma}^\dagger c_{d1\sigma} + c_{r\sigma}^\dagger c_{d2\sigma}  + h.c.].
\label{Hdotsleads} 
\end{equation}

\subsection{Conductance Calculation}
The current at any time $t$ between nearest-neighbors sites $i$ and $j$ is calculated as
\begin {equation}
J_{ij}(t) = i{2\pi e\over h}t_{ij}\sum_{\sigma} \langle \Psi(t)|(c^{\dagger}_{i\sigma} c_{j\sigma} - c^{\dagger}_{j\sigma} 
c_{i\sigma})| \Psi(t)\rangle, 
\label{Jij}
\end {equation}
where $|\Psi(t)\rangle$ is the wave function of the system at time $t$, which will be calculated with the
DMRG method, using a number $M$ of states in the process.  $c_{i\sigma}^\dagger$ creates an electron 
with spin $\sigma$ at site $i$, which can be part of the interacting region or be at the leads.
In the results presented below, the current shown without any link or site index corresponds to 
\begin {equation}
J(t) = (J_L(t) + J_R(t))/2,
\label{Jij2}
\end {equation}
where $J_L(t)$ is the current between the last site of the left lead and the first dot, 
and $J_R(t)$ is the current between 
the last dot and the first site of the right lead, moving from left to right.
The conductance $G$ can be obtained by simply dividing the steady-state current by the total bias $\Delta V$.
The individual voltages $\pm$$\Delta V$/2 in the leads are applied uniformly in each one, as indicated in Fig.\ref{schematic}.
Note that the use of a symmetrized current $J(t)$ is not crucial, since a good left-right symmetry
is observed when currents at several links are studied, as described in more detail below.

\subsection{Technique}
Closely following Ref.\onlinecite{White}, a brief description of the numerical technique is here provided for
completeness.  The basic idea is to incorporate the Suzuki-Trotter (ST) decomposition of the time-evolution 
operator \cite{Vidal} into the DMRG finite-system algorithm \cite{White,Daley}.  
The second order ST decomposition of the 1D Hamiltonian as employed in Ref.\onlinecite{White}
can be written as 
\begin {equation}
e^{-i\tau H}\approx e^{-i\tau H_1/2} e^{-i\tau H_2/2}...e^{-i\tau H_2/2}e^{-i\tau H_1/2},
\label{ST}
\end {equation} 
where $H_j$ is the Hamiltonian of the link $j$.
The DMRG representation of the wavefunction at a particular step $j$ during the finite-system sweep is 
\begin {equation}
|\psi\rangle = \sum_{l\alpha_j\alpha_{j+1} r} \psi_{l\alpha_j\alpha_{j+1} r}|l\rangle|\alpha_j\rangle|\alpha_{j+1}\rangle|r\rangle,
\label{DMRG}
\end {equation}
where $l$ and $r$ represent the states of the left and right blocks (in a truncated basis, optimally selected as eigenvectors of 
a density matrix), 
while $\alpha_i$ and $\alpha_{j+1}$ represent the states of 
the two central sites.  An operator $A$ acting on sites $j$ and $j+1$ (namely, only involving nearest-neighbors) 
can be applied to $|\psi\rangle$ exactly, and re-expressed 
in the same optimal basis as
\begin {equation} 
[A\psi]_{l\alpha_j\alpha_{j+1} r} = \sum_{\alpha'_j\alpha'_{j+1}} A_{\alpha_j\alpha_{j+1};\alpha'_j\alpha'_{j+1}}
\psi_{l\alpha'_j\alpha'_{j+1} r}.
\label{Operator}
\end{equation}
Thus, the time evolution operator of the link $j$ can be applied exactly on the DMRG step $j$.  As a consequence, 
the time evolution 
is done by applying $e^{-i\tau H_1/2}$ at DMRG step 1, $e^{-i\tau H_2/2}$ at DMRG step 2, and so on, thus forming the usual 
left-to-right sweep. Then, applying all the reverse terms in the right-to-left sweep.  A full sweep evolves the system  
one time step $\tau$.  The error introduced by the second order decomposition is order $\tau^3$ 
in each time step \cite{White}.  Thus, upon 
evolving the system one time unit ($1/\tau$ steps), an order $\tau^2$ error is introduced. Numerically,
the influence of this small systematic error is easy to control. This brief summary gives the reader a rough
idea of the technique used here. Details regarding lattice sizes, number of states kept in the DMRG procedure, and influence
of other parameters are discussed below.

\section{NON-INTERACTING CASE}

\begin{figure}
\vspace{0.5cm}
\epsfxsize=8cm \centerline{\epsfbox{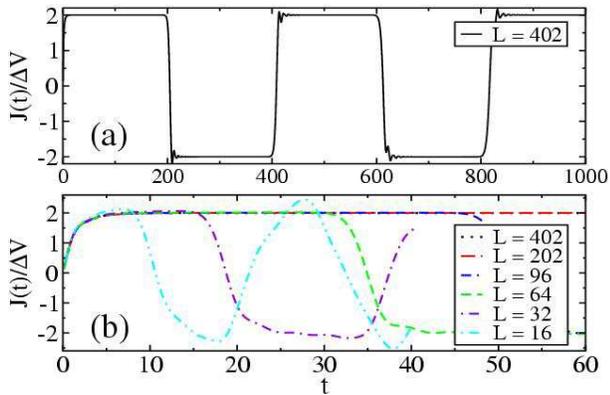}}
\caption{Exact results for $J(t)/\Delta V$ (in units of $e^2/h$) vs. time 
(in units of $\hbar$/$t_{\rm leads}$),
for the non-interacting 1QD case obtained with clusters 
of different lengths ($L$) and $\Delta V$=0.001.  
(a) $J(t)/\Delta V$ obtained with a large cluster ($L$=$402$).  
$J(t)/\Delta V$ shows clear 
steady-state plateaus at $\pm 2e^2/h$.  The periodic changes in the current direction are caused by its reflection at the 
open boundaries of the cluster.  (b) $J(t)/\Delta V$ obtained with decreasing $L$.  The steady-state plateau is obtained 
even with $L$ = $32$.  The current is quasi-periodic with a period proportional to $L$.  The parameters used 
are $V_{\rm g}$=$U$=$0$ 
and $t'$=$0.4$.}
\label{exact.L}
\vspace{0.25cm}
\end{figure}

Properties of the method discussed in this paper are exemplified
in Fig.\ref{exact.L}(a), where the current at the center of the chain is
shown (divided by the voltage difference) as a function of time. 
The current in this figure is {\it exactly}
calculated, not using DMRG, since for non-interacting particles the problem reduces to a single
electron problem. A small bias
$\Delta V$=0.001 will be used, unless otherwise stated. Thus, in this first
study the focus will be on trying to reproduce results expected from linear response,
but a few results with a finite bias are also included in the manuscript, as discussed below.
Returning to Fig.\ref{exact.L}(a), for a bulk system the current would be expected to raise for a small fraction
of time, and then reach a steady state. This indeed occurs even in our
finite-size systems. In fact, the transition from zero current at $t$=0
to an approximately time-independent current regime is very fast, 
and can be barely observed
in the scale of Fig.\ref{exact.L}(a). But the existence of a very flat plateau in
the current is clear, and its value will be used to extract the conductance
below. Note that due to the finite size of the system, the current cannot continue in the same
steady state at all times. The Hamiltonian is particle number conserving and, as a consequence,
the presence of a current implies a population/depopulation of the leads, which cannot continue forever.
In fact, once the front of the charge pulse reaches the end of the chain,
it bounces back and eventually generates a current of the opposite sign. This effect
will be discussed in more detail later. Here, it is important to remark that in spite
of this periodicity present in finite open-end systems, 
the flat plateaus are clearly defined over an extended period of time for the lattice of 402 sites used,
and the value of the conductance can be easily deduced from those individual plateaus,
as discussed below. Note that the setup of Fig.\ref{schematic} 
and the existence of plateaus in the current Fig.\ref{exact.L}
are natural in the DMRG/transport
context and was observed before \cite{peter}. 
Our main contribution will be the use of the adaptive DMRG
method for the calculations, as shown below.

\begin{figure}
\vspace{0.5cm}
\epsfxsize=9cm \centerline{\epsfbox{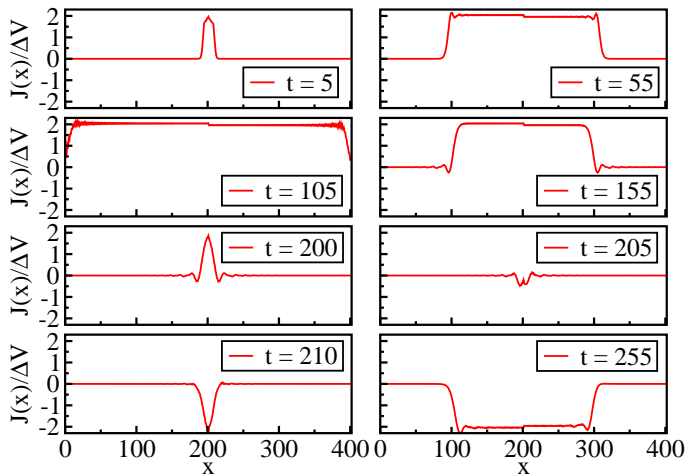}}
\caption{This figure shows the propagation of the current in the cluster after the bias $\Delta V$=0.001 is applied at $t$=0.  
The different panels show the current as a function of position $x$ 
along the chain, at different times $t$.  The size is $L = 402$ and 
the dot is at the site 201. The results were obtained exactly, since the Hubbard couplings are zero.}
\label{G.x}
\vspace{0.25cm}
\end{figure}

To further illustrate the propagation of charge in the cluster after the finite bias is switched on at time $t$=0,
in Fig.\ref{G.x} the exact current at different positions $x$ is shown, parametric with time. At small times, $t$=5 (in
units of $\hbar$/$t$) only the central portion is affected as expected. At time $t$=55, the affected region is much
larger, while at $t$=105, the front has reached the ends of the chain and soon after it starts bouncing back. At
times $t$=200 and 205, approximately the initial condition is recovered, and almost everywhere the current to the left
and right cancel out nearly exactly. For larger times, a reverse sign current is created. Note that in our studies
there are $no$ sources of dissipation, and the current will keep on oscillating forever. Adding inelastic processes
is a next major challenge in this context, left for the future.

\begin{figure}
\vspace{0.5cm}
\epsfxsize=8cm \centerline{\epsfbox{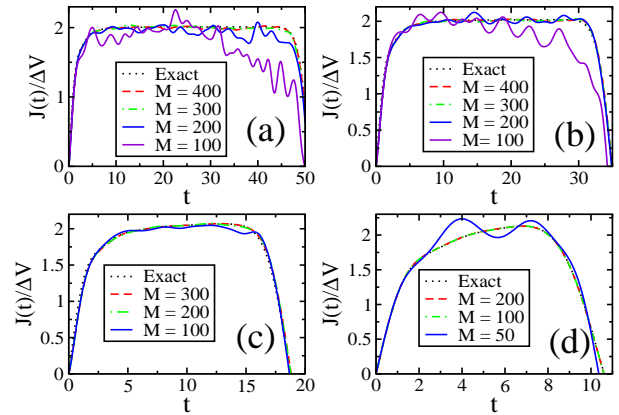}}
\caption{DMRG results compared to the exact results for $J(t)/\Delta V$ obtained using different clusters $L$ and number of 
states $M$, with $\Delta V$=0.001. 
(a) $L$ = 96, (b) $L$ = 64, (c) $L$ = 32, and (d) $L$ = 16.  Note that for $L$ = 96 and 64, $M$ = 200 shows good qualitative 
agreement and $M$ $\ge$ 300 even shows good quantitative agreement with the exact results.  For $L$ = 32 and 16, $M$ = 200 and 
100 already show excellent quantitative agreement with the exact results.}
\label{DMRG-Exact.L}
\vspace{0.25cm}
\end{figure}

It is also important to show that the existence of the plateaus is not restricted to
very long chains of hundreds of sites, but they are visible on much smaller systems,
increasing the chances that the numerical DMRG method can be used even for complicated nanosystems.
Figure \ref{exact.L}(b) contains the current vs. time for a variety of lattice sizes,
ranging from 402 to systems as small as 16 sites. The time width of the plateaus 
depends on $L$, as expected, but the value of the current at the plateaus is approximately
$L$ independent
even up to systems as small as containing $L$=32 sites. Even the $L$=16 chain has a periodicity with
a first plateau in the current which is also in good agreement with the expected value from larger sizes.
Thus, this behavior appears to be robust and 
the plateaus are also expected to be present for the chains that the DMRG method can handle. 
That this is the case can be shown in Fig.\ref{DMRG-Exact.L},
where DMRG results for the current vs. time are shown, compared with exact data. Consider first a sufficiently
long chain, as shown in (a), such that a sharp plateau is observed in the exact result.
Figure \ref{DMRG-Exact.L}(a) shows that increasing the number of states $M$ used in the DMRG approximation, a 
convergence to the exact solution is observed. In fact, for $M$=300 or higher,
the DMRG results cannot be distinguished from the exact ones. A similar behavior
is found using shorter chains as for the case with $L$=64 sites Fig.\ref{DMRG-Exact.L}(b),
but in this example the plateau can be observed accurately even with a smaller number of states
such as $M$=200. The trend continues for smaller systems (c,d): For $L$=32 and 16,
the DMRG method reproduces the exact results with high accuracy using $\sim$100
states.

\begin{figure}
\vspace{0.5cm}
\epsfxsize=8cm \centerline{\epsfbox{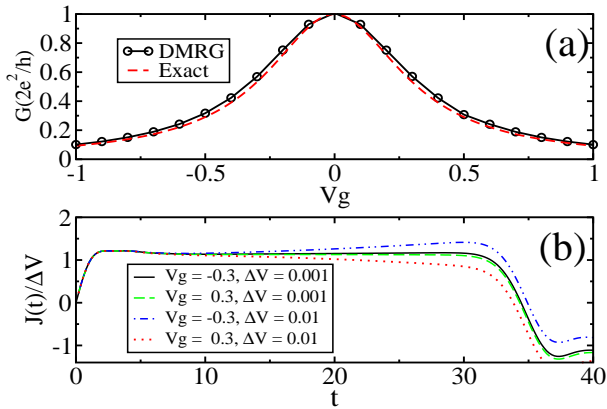}}
\caption{(a) DMRG and exact results for $G$ vs. $V_{\rm g}$ in the case of one non-interacting QD, and $t'$ = 0.4 (with $\Delta V$=0.001).  The DMRG 
results are obtained with $L$ = 64 and $M$ = 300.  In this case, $G$ is obtained from the value of the steady state current plateau. 
The exact results are for infinite leads. The results show a resonant tunneling peak of FWHM = $4 t'^2$ at $V_{\rm g}$ = 0. (b)  
$J(t)/\Delta V$ for $V_{\rm g}$ = $\pm$0.3 and $\Delta V$ = 0.01, 0.001.  Note that decreasing the bias voltage reduces the asymmetry 
between positive and negative $V_{\rm g}$ and gives a better steady-state plateau.}
\label{Exact.G-Vg}
\vspace{0.25cm}
\end{figure}

Figure \ref{Exact.G-Vg}(a) 
shows the conductance deduced from the behavior of the
current obtained with the DMRG method, versus the gate voltage $V_{\rm g}$, for
the case of a single ``noninteracting'' quantum dot, namely one having $U$=0. The hopping
amplitude between the dot and the leads is $t'$=0.4. It is expected that the
maximum value of the conductance be obtained when the level in the dot is
aligned with the Fermi level of the leads, and this occurs in our case at $V_{\rm g}$=0.
The DMRG results beautifully confirm this expectation. As the gate voltage changes
away from 0, the conductance is expected to decrease symmetrically and this is
indeed shown in Fig.\ref{Exact.G-Vg}(a). In fact, the results at nonzero gate voltage
are also in excellent quantitative agreement with the exact results.

All the previous results were obtained for a sufficiently small value of the bias voltage
$\Delta$$V$=0.001,
as already explained. It is interesting to observe how the results change
when larger values of $\Delta$$V$ are employed.
Figure \ref{Exact.G-Vg}(b) shows the current for a couple of biases. The existence
of the plateau is clear in both cases, but for $\Delta$$V$=0.01, asymmetries between
positive and negative bias can be observed, which are not expected in the limit
$\Delta$$V$$\rightarrow$0. As a consequence, the rest of the results discussed
below were obtained with $\Delta$$V$=0.001, unless otherwise noted. 
An easy criterion to realize if a sufficiently small bias is used to obtain
the linear response limit is to repeat the calculations for the same amplitude of
bias, but opposite signs, and see if a noticeable difference is obtained.

\begin{figure}
\vspace{0.5cm}
\epsfxsize=7cm \centerline{\epsfbox{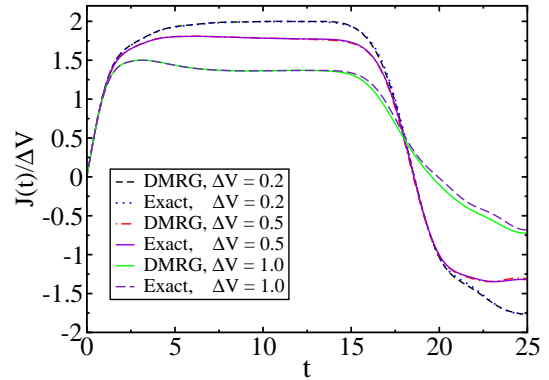}}
\caption{(a) DMRG results compared to the exact results in the case of one non-interacting QD for several intermediate and 
large values of $\Delta V$, namely exploring the influence of a finite bias in the calculations.
 The parameters used are $V_{\rm g}$ = $U$ = 0, and $t'$ = 0.4.   Both DMRG and exact results are obtained with 
a cluster $L$ = 32, using $M$ = 200 states, for the DMRG results.}
\label{DMRG-Exact.dV}
\vspace{0.25cm}
\end{figure}

The method proposed here also works in the case of a finite bias voltage, namely
it is not restricted to the linear response regime. To show that the technique can handle
even a large bias, in Fig.\ref{DMRG-Exact.dV} results for the current vs. time
are shown at the indicated voltages. It is only at $\Delta$$V$ as large as 1.0
that small differences are visually observed in the figure, between the DMRG and
exact results. This can be easily fixed increasing slightly the number of states $M$.
Thus, overall the method appears to be sufficiently robust to handle arbitrary
voltages, showing the generality of the technique here proposed. Nevertheless,
further work in the finite bias context 
will be important to fully test this case, calculating
differential conductances and analyzing the regime of very strong bias.

\begin{figure}
\vspace{0.5cm}
\epsfxsize=8cm \centerline{\epsfbox{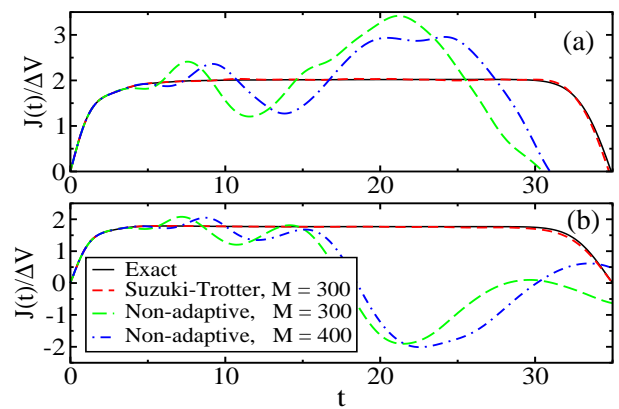}}
\caption{The results obtained with the Suzuki-Trotter approach and the static Runge-Kutta method  compared to the exact results 
for (a) $\Delta V = 0.001$ and (b) $\Delta V = 0.5$.  The parameters used are $V_{\rm g}$ = $U$ = 0.0, $t'$ = 0.4, and $L$ = 64.}
\label{Methods-comp}
\vspace{0.25cm}
\end{figure}

In our investigations, the numerical study was also carried out using
a ``static'' procedure, where the $t$=0 DMRG basis is not expanded
with growing time. In this case, the results are obtained by integrating the 
time-dependent Schr\"odinger equation using the fourth order Runge-Kutta method, and also
using the DMRG ground state as the initial state.\cite{marston} This is to be contrasted
with the procedure of Refs.\onlinecite{White,Daley} where the basis is modified with time. Figure \ref{Methods-comp}
shows the results of both procedures: clearly using an adaptive basis provides superior data,
reproducing accurately the exact results. In the static procedure 
a similar accuracy is reached only by increasing substantially the number of states, thus
missing its economical CPU-time advantages. 

\begin{figure}
\vspace{0.5cm}
\epsfxsize=8cm \centerline{\epsfbox{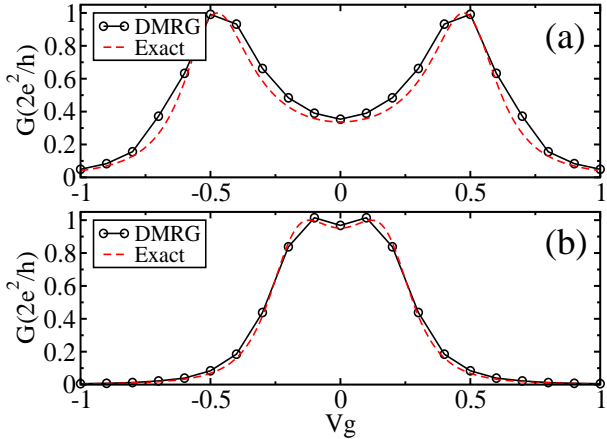}}
\caption{DMRG and exact results for $G$ vs. $V_{\rm g}$ in the case of 2 coupled non-interacting QD's.  The DMRG results 
are obtained with $L$ = 64 and $M$ = 300 ($\Delta V$=0.001).  
The exact results are for infinite leads.  The cases (a) $t''$ = 0.5 and 
(b) $t''$ = 0.2 are investigated, with
$t'$ = 0.4 in both cases.  The results present the bonding and anti-bonding resonant tunneling peaks at $\pm t''$.}
\label{2QD.G-Vg}
\vspace{0.25cm}
\end{figure}

The method discussed here also works nicely for the case of two non-interacting QD's,
as shown in Fig.\ref{2QD.G-Vg} for two different values of the hopping amplitude
$t''$ between the dots. The slight difference between the DMRG and the bulk exact results
can be improved increasing the number of sites.

\begin{figure}
\vspace{0.5cm}
\epsfxsize=8cm \centerline{\epsfbox{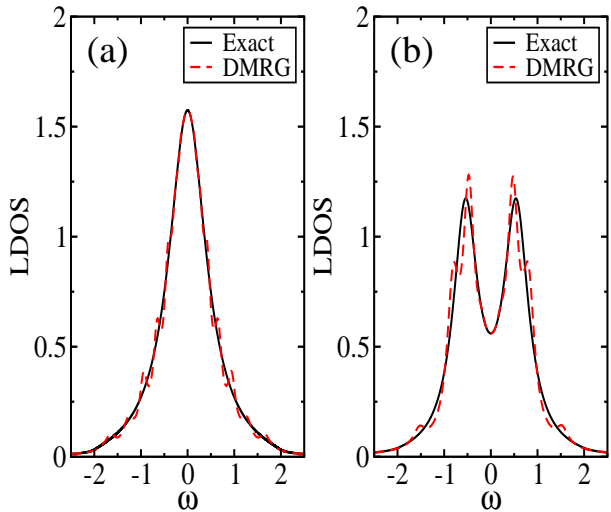}}
\caption{DMRG local density-of-states results for (a) 1QD and (b) 2QD's 
compared to the exact (bulk) results.  The parameters used 
are $V_{\rm g}$ = $U$ = 0, $t'$ = 0.4 in both cases and $t''$ = 0.5 in (b).  In (a), a broadening 
imaginary component $\eta = 0.1$ was introduced.  In (b), $\eta = 0.15$.  Smaller values of $\eta$ 
would reveal the discrete nature of the LDOS obtained with DMRG on a finite $L$=64 system.}
\label{LDOS}
\vspace{0.25cm}
\end{figure}

Although not directly related with the method to obtain
the conductance of an interacting nano-system discussed in this paper, for completeness we have also studied
the local density-of-states which are important to guide the intuition and contrast with
other methods and scanning tunneling microscopy experiments as well. In Fig.\ref{LDOS}, results for non-interacting quantum
dots are shown (namely, dots where the Hubbard repulsion is 0). Clearly, both the
exact results (which are shown already in the bulk limit)
and the DMRG results, slightly smeared by shifting, using a small imaginary component $\eta$, the
pole locations in the continued fraction expansion, are in excellent agreement in both cases. 
A smaller $\eta$ would have revealed the many $\delta$-functions in the DMRG case obtained
using a finite chain with 64 sites.

\section{ONE QUANTUM DOT}

In the previous sections, the method was introduced and tested for the
case of noninteracting $U$=0 electrons. But the main application of the
technique is envisioned to occur in the presence of nontrivial Coulombic
interactions (and eventually also adding phononic degrees of freedom). 
In this section, the case of a nonzero Hubbard coupling will
be considered, focusing on the special case of one quantum dot. The Hamiltonian
used was already discussed in previous sections.

\subsection{Results at intermediate values of U}
\begin{figure}
\vspace{0.5cm}
\epsfxsize=6.5cm \centerline{\epsfbox{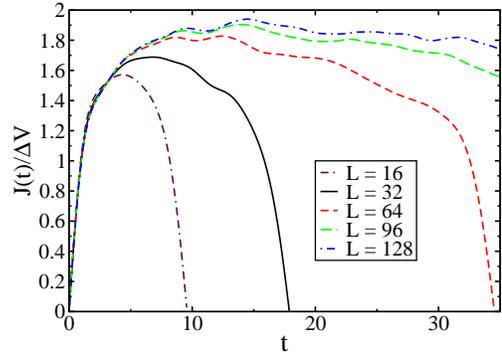}}
\caption{$J(t)/\Delta V$ in the case of one quantum dot for different cluster lengths $L$. The parameters 
used are $U$ = 1.0, $t'$ = 0.4, $\Delta V$=0.001, 
and $V_{\rm g}$ = -0.5.  As $L$ increases, the conductance approaches the unitarity limit ($2e^2/h$) 
due to the Kondo screening effect.}
\label{U.1.0.L}
\vspace{0.25cm}
\end{figure}

Figure \ref{U.1.0.L} contains our DMRG results for the current vs. time, for
the case of $U$=1.0. Similar values of this coupling were extensively used in previous
investigations \cite{ED+DE1,Busser1,Busser2,Martins1,Martins2,Khaled}, 
and it is believed to lead to a Kondo cloud of a size amenable
to numerical investigations (note that if $U$ is very large, the effective $J$ between localized
and mobile spins is reduced,
and it is known that the cloud's size rapidly grows with decreasing $J$). The figure shows that the systematic
behavior found in the noninteracting case survives the presence of a Coulomb
interaction, namely the current develops plateaus that can be used to determine
the conductance. For instance, this effect is clearly present for $L$=96 and 128, although
for smaller sizes (shown for completeness) the maximum current is 10\% to 20\% less
than expected and one must be cautious with size effects. The value of
the gate voltage is -$U$/2, which in the absence of the Kondo effect would locate the system
in the conductance ``valley'' (implying a near zero conductance) between the Coulomb blockade peaks at -$U$ and 0. 
The figure shows that the method
introduced in this paper is able to reproduce the existence of a Kondo effect,
since the conductance is actually very close to the ideal limit 2$e^2$/$h$ \cite{Kondo-theory,ED+DE1,Busser1,Busser2}, 
rather than being negligible. This is a highly
nontrivial test that the proposed technique has passed.

\begin{figure}
\vspace{0.5cm}
\epsfxsize=6.5cm \centerline{\epsfbox{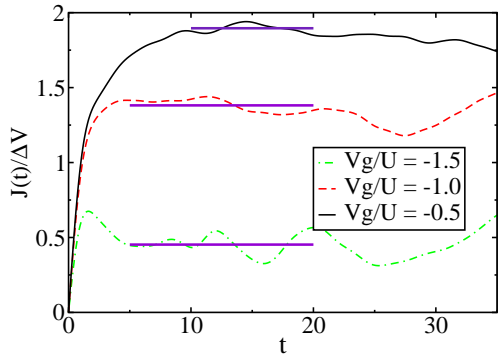}}
\caption{$J(t)/\Delta V$ in the case of one interacting QD for different values of $V_{\rm g}$, 
and with $\Delta V$=0.001. The value of the conductance 
is obtained by averaging the current over an interval of time, corresponding to the steady state.  
The solid horizontal 
lines represent this time interval over which the average of the current is taken, and the value of the average.  The 
parameters used are $U$ = 1.0, $t'$ = 0.4, $L$ = 128, and $M$ = 350.}
\label{average}
\vspace{0.25cm}
\end{figure}

Results for other values of the gate voltage are shown in Fig.\ref{average}. Moving
away from $V_{\rm g}$=-$U$/2, the current is reduced, as discussed below in more detail.
Note that the plateaus contain small oscillations as a function of time.\cite{wingreen} The procedure
followed here to extract the current needed for the calculation of the conductance is to
carry out averages over time, as shown in the figure, once the plateau region is reached.
The size of the oscillations give an indication of the errors in the numerical determination
of the conductance, for a given lattice size. As $L$ and $M$ increases, the oscillations tend to
disappear.

\begin{figure}
\vspace{0.5cm}
\epsfxsize=8cm \centerline{\epsfbox{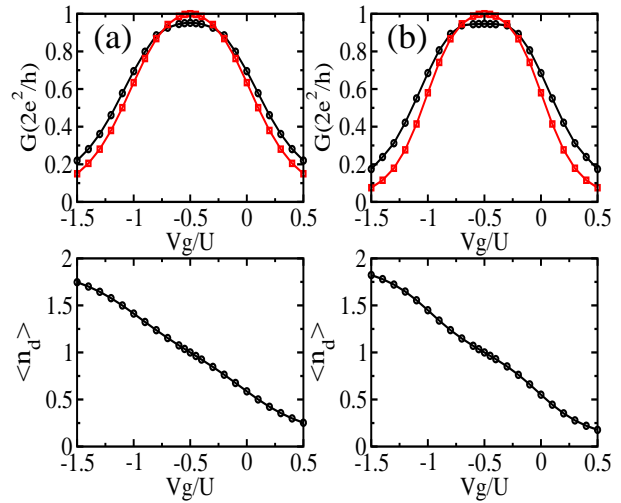}}
\caption{Conductance $G$ ($\Delta V$=0.001)
and the dot occupation $\langle n_d \rangle$ for one interacting QD.  The 
circles show $G$ obtained by averaging the current over an interval of time corresponding to the 
steady state, as shown in Fig.\ref{average}.  The squares show $G$ obtained from $\langle n_d \rangle$ 
using the Friedel sum rule (FSR).  $G$ has the shape of the expected Kondo or mixed-valence plateau 
centered at $V_{\rm g}$ = $-U/2$. The 
feature would be sharper reducing $t'$. Results shown are: (a) $U$ = 1.0, $t'$ = 0.4, and  (b) $U$ = 
2.0, $t'$ = 0.5. In both cases $L$ = 128 and $M$ = 350.  Note that the DMRG conductance
results in (a) show a slightly better 
agreement with the FSR results. This is expected since the finite size effects are stronger for larger $U$.}
\label{G.N-Vg}
\vspace{0.25cm}
\end{figure}

Following the procedure sketched in Fig.\ref{average}, the full conductance vs. $V_{\rm g}$
was prepared for $U$ equal to 1 and 2. The results are in Fig.\ref{G.N-Vg}. The shape of the
curve is the expected one for the regime considered here: the intermediate values of $U$ do not
locate our investigation deep in the Kondo regime, with sharply defined integer charge at the dot,
but more into the mixed-valence region. This can be deduced from the value of the dot charge 
vs. $V_{\rm g}$, also shown in Fig.\ref{G.N-Vg}. With increasing $U$, sharper charge steps
are formed, but the Kondo cloud size increases, as discussed later.

\subsection{Results at large values of U}
\begin{figure}
\vspace{0.5cm}
\epsfxsize=8cm \centerline{\epsfbox{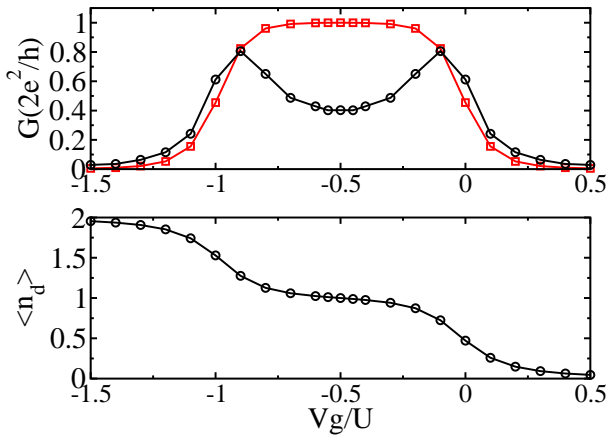}}
\caption{Conductance $G$ ($\Delta V$=0.001) and charge at the dot $\langle n_d \rangle$, 
for one interacting QD in the case of large $U$.  
The circles show $G$ obtained from DMRG current procedure outlines in this paper, while
the squares show $G$ obtained from $\langle n_d \rangle$ 
using the Friedel sum-rule.  The finite-size effects are obvious here, since 
the results are halfway between the expected Kondo 
plateau (properly reproduced by the FSR procedure) 
and the Coulomb blockade peaks. This case is shown as an illustration of important size effects in some limits.
The parameters used are $U$ = $4.0$, $t'$ = $0.4$, $L$ = $128$, and  
$M$ = $300$. 
}
\label{G.N-Vg.U.4.0}
\vspace{0.25cm}
\end{figure}

There are cases where the technique gives results that are only qualitatively correct. While clearly 
further increases of the number of states and lattice sizes will improve the accuracy, it is important
to judge if at least the essence of the physics has been captured by our proposed method. 
In Fig.\ref{G.N-Vg.U.4.0}, results for $U$=4 are shown. This is a representative of the ``large'' $U$
regime, since it must be compared with $t'$ (as opposed to $t$=1) that is only 0.4 in this figure.
Another indication that this $U$ is 
large is in Fig.\ref{G.N-Vg.U.4.0}(b) where a clear sharp quantization
of the charge inside the dot is observed. In this large-$U$ regime, the DMRG conductance is shown in (a).
Clearly, there is a substantial difference between the Friedel sum-rule estimation (which has the correct
``box'' shape in the gate voltage range -1 and 0) and the DMRG numbers. However, at least the fact that
there must be a nonzero conductance at $V_{\rm g}$=-$U$/2 was properly captured by the method. This example
illustrates a case where size effects are important, due to the subtle rapid increase of the Kondo cloud
with increasing $U$. However, the qualitative results were captured, in spite
of the fact that
to reach a quantitative conclusion much
larger sizes must be considered. This case is shown as a cautionary example to the readers, that must be
alert of the 
limitations of the numerical methods. Note that while FSR results are excellent, for other arbitrary cases it would be
unclear whether the Friedel sum-rule method is valid and, as a consequence, not always this procedure can be used.

\subsection{Improving the Convergence}
\begin{figure}
\vspace{0.5cm}
\epsfxsize=6cm \centerline{\epsfbox{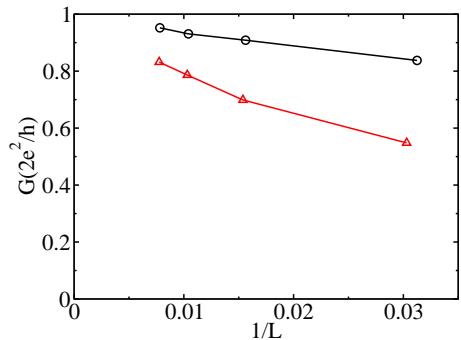}}
\caption{ Finite-size scaling of the conductance $G$ at $V_{\rm g}$ = -$U/2$ for the odd-1QD-even  
(circles) and even-1QD-even clusters (triangles) setups.  Note that in both cases $G$ converges to 1 in the bulk limit. 
However, the odd-1QD-even cluster converges faster, which makes it the most useful
for practical calculations.  The parameters used are $U$ = 1.0 and $t'$ = 0.4.}
\label{FiniteSize}
\vspace{0.25cm}
\end{figure}

To carry out the investigations presented thus far in this and the previous section, the number of
sites in the leads at left and right of the dot region have been chosen such that one lead has
an $even$ number and the other an $odd$ number of sites (to refer to this case, the notation used below will 
be ``odd-QDs-even''). While this should be irrelevant for very long chains, 
in practice this is important for the speed of convergence
of the conductance calculations with increasing cluster size. For example, in the interacting case $U$$\neq$0,
a Kondo or mixed-valence regime is expected where the spin at the dot couples with electrons
at the Fermi level of the leads. This formation of the Kondo cloud occurs more efficiently on
a finite-size lattice if already a zero energy level is available, as it occurs when one of the leads
has an odd number of sites. That this improves the rate of convergence with increasing lattice
size is shown in Fig.\ref{FiniteSize}, where the odd-1QD-even case is contrasted to the 
even-1QD-even case, where in both leads the number of sites is even. Clearly, the odd-1QD-even
case approaches the ideal limit 2$e^2$/$h$ faster than using leads with an even number of sites,
and it is recommended to be used in future investigations. The figure also shows that eventually with
sufficiently large systems, both combinations will reach the same ideal result, as expected.
The remaining possibility (odd-1QD-odd) was also investigated (not shown). 
For reasons that remain to be analyzed further, the convergence in
this case is not as good as in the odd-1QD-even case.
As a consequence, empirically it is clear that the combination even-QDs-odd is the most
optimal to speed up the size convergence of the calculation. 

\begin{figure}
\vspace{0.5cm}
\epsfxsize=6cm \centerline{\epsfbox{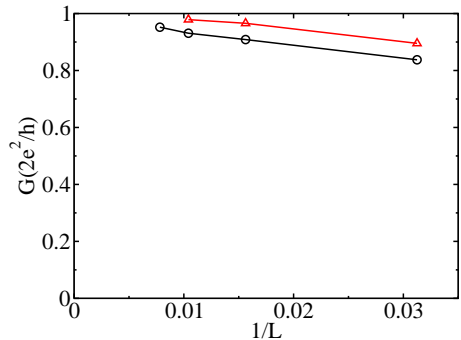}}
\caption{Finite size scaling of $G$ for odd-QD-even clusters using OBC (circles) and DBC (triangles).  
$d = 0.5$ was used.  Note that $L=64$ cluster with DBC gives better results than a $L$=$128$ cluster 
with OBC.  The same parameters are used as in Fig.\ref{FiniteSize}.}
\label{FiniteSize2}
\vspace{0.25cm}
\end{figure}

Another method to improve the size convergence was tested.
Following Ref.~\onlinecite{Schmitteckert2}, the finite-size effects can be reduced by 
using ``damped boundary conditions'' (DBC).  The hoppings in the $M_D$ links at the boundaries  
are reduced using the formulas $-td, -td^2,..., -td^{M_D}$, where $d$$<$$1$.  $M_D$ has to be chosen such that the 
damping occurs far enough from the central region.  Figure \ref{FiniteSize2} shows the 
finite size scaling using odd-QD-even clusters and the same parameters as in Fig.\ref{FiniteSize}, 
both with DBC and the standard open boundary conditions (OBC) used in the rest of the paper.
The latter indeed improves the convergence.
While in the present study, OBC were used in most of the manuscript to keep the simplicity in the presentation
and reduce the number of parameters, the use of DBC is recommended for cases where size effects are strong. 

\subsection{Influence of Magnetic Fields}
\begin{figure}
\vspace{0.5cm}
\epsfxsize=6.5cm \centerline{\epsfbox{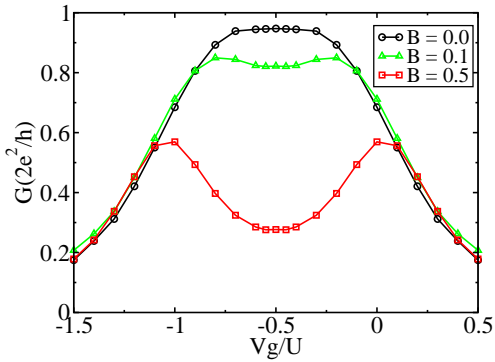}}
\caption{Conductance $G$ vs. gate voltage $V_{\rm g}$ 
in the case of one interacting QD ($U$ = 2.0, $t'$ = 0.5) for different values of the magnetic field $B$
(and $\Delta V$=0.001).  For 
$B$ = 0, a Kondo plateau is obtained, centered at $V_{\rm g}$ = $-U/2$.  
As $B$ increases, the Kondo effect is suppressed, and 
for moderate $B$, two Coulomb blockade peaks are observed at $V_{\rm g}$ = $-U$ and $V_{\rm g}$ = 0, as expected.}
\label{G-Vg.B}
\vspace{0.25cm}
\end{figure}

To fully confirm that our investigations in the Kondo/mixed valence regime have captured
the essence of the problem, namely the formation of a Kondo cloud with antiferromagnetic
coupling between the spins at the leads and the dot, investigations including magnetic
fields are necessary. In Fig.\ref{G-Vg.B}, it is shown how the the Kondo plateau in the conductance 
evolves with increasing magnetic field. As expected from previous investigations, including results obtained 
using very different techniques such as the Lanczos method followed by a Dyson-equation 
embedding procedure (ED+DE) \cite {Torio}, the conductance broad peak splits with increasing magnetic field $B$.
At large $B$, two peaks are observed at -$U$ and 0, as it occurs also in the high temperature
regime where only Coulomb blockade effects are present.

\section{TWO COUPLED QUANTUM DOTS}

\begin{figure}
\vspace{0.5cm}
\epsfxsize=7cm \centerline{\epsfbox{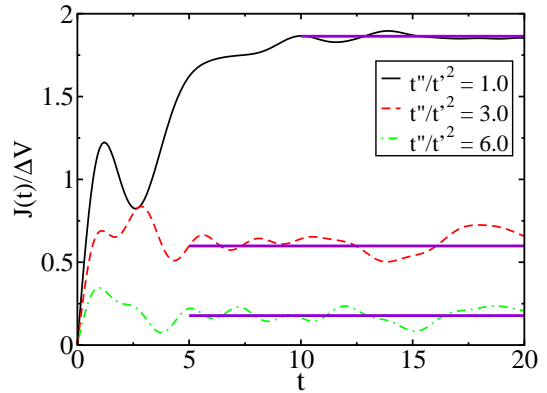}}
\caption{$J(t)/\Delta V$ for two coupled QD's at $V_{\rm g}$ = $-U/2$ for different 
values of $t''/t'^2$ (and $\Delta V$=0.001).  As in the case of one 
interacting QD, the conductance is obtained by averaging the steady-state current over the indicated intervals.  The parameters 
used are $U$ = 1.0, $t'$ = 0.5, $L$ = 127, and $M$ = 350.}
\label{2QD}
\vspace{0.25cm}
\end{figure}

The method discussed in this paper is general, and in principle it can be implemented for a variety
of complicated geometries and couplings in the interactive region between the leads. Thus, it is
important to confirm that the method will keep its validity going beyond the one quantum dot
case. In this section, the case of two dots will be studied. Systems with two quantum  dots
are believed to be understood theoretically and, as a consequence, our numerical data can be contrasted
against robust results in the literature. Cases involving more dots \cite{Busser1} are still not fully
understood, and their analysis will be postponed for future investigations.
In Fig.\ref{2QD}, the current vs. time is shown for two dots. The Hamiltonian
for this case was already defined in previous sections. For a fixed $t'$, increasing the amplitude of the
direct hopping between the dots $t''$ amounts
to isolating the two dots system from the rest. As a consequence, the current is expected to
decrease, and the method indeed reproduces this effect, as
shown in the figure. The same physics is obtained reducing $t'$, at fixed $t''$. In fact, previous studies \cite{Georges}
have shown that the conductance only depends on $t''$/$t'^2$, and this has been verified
using our method.

\begin{figure}
\vspace{1.0cm}
\epsfxsize=6cm \centerline{\epsfbox{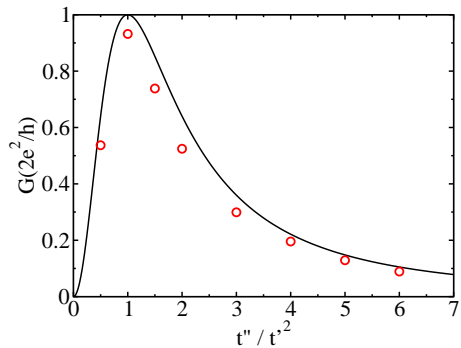}}
\caption{Conductance $G$ 
as a function of $t''/t'^2$, at $V_{\rm g}$ = $-U/2$ ($\Delta V$=0.001). 
In this regime, $G$ is determined by the competition between the Kondo 
correlation of each dot with the neighboring leads and the antiferromagnetic correlation between the two dots.  The circles 
represent our DMRG results obtained with $L$ = 127 and $M$ = 350.  The solid line is the plot of the functional form obtained 
by Georges and Meir using SBMFT.\cite{Georges}}
\label{G-tpp}
\vspace{0.25cm}
\end{figure}

The conductance of a system with two dots in series will decrease with increasing
$t''$/$t'^2$ (at large $t''$/$t'^2$)
due to the decoupling of the two-dots system into a small two-sites molecule, 
as already discussed. But this conductance will also be very small at small $t''$ when the
tunneling from one dot to the next is nearly cutoff. As a consequence, the conductance vs. $t''$/$t'^2$
is known to present a peak at intermediate values. In Fig.\ref{G-tpp}, the slave-boson mean-field technique (SBMFT) 
predictions for this case obtained in previous investigations \cite{Georges} are shown together with our results.
The agreement is fairly reasonable, providing further support that the method discussed here can handle
systems where there are competing tendencies, beyond the one quantum-dot case.

\section{Conclusions}

In this manuscript, a method was proposed and tested to calculate the conductance of small (nanoscale) strongly
correlated systems modeled by tight-binding Hamiltonians. The approach is based on the adaptive time-dependent
DMRG method, and it was shown to work properly for non-interacting systems and also in the cases of one and two
quantum dots. Besides the finite size effects, discussed in the text as well, there are no other severe limitations
to handle complex interacting models with arbitrary couplings. The method is a complement to DFT calculations
in the nanoscopic context. Further improvements of the technique must consider temperature and inelastic effects.

It is concluded that the semi-quantitative
analysis of transport in models of strongly correlated nanosystems appears under reach, and much
progress is expected from the application of the technique presented here to
realistic Hamiltonians for small molecules and arrays of quantum dots. The remarkable
cross-fertilization between modeling, simulation, 
and experiments existing in bulk strongly correlated materials, such
as transition-metal oxides, can be repeated in a variety of interesting systems at the nanoscale.

\section{Acknowledgments}

The authors thank Brad Marston,
Eduardo Mucciolo, Peter Schmitteckert, and Uli Schollw\"ock for useful comments.
The work of K.A-H., C.B., and E.D. was partially supported by
the NSF Grant No. DMR-0454504.


\end{document}